\documentclass{iopconfser}
\usepackage{subcaption}
\begin{document}

\title{Classification and stability of black hole event horizon births: a contact geometry approach}

\author{Oscar Meneses-Rojas}

\affil{Universit\'e Bourgogne Europe, UMR 5584, CNRS, F-21000 Dijon, France}
\email{oscar.meneses-rojas@univ-lille.fr}

\begin{abstract}
A classical result by Penrose establishes that null geodesics generating a black hole event horizon can only intersect at their entrance to the horizon in ``crossover'' points. This points together with limit points of this set, namely caustics, form the so-called "crease set". Light rays enter into the horizon through the crease set, characterizing the latter as the birth of the horizon. A natural question in this context refers to the classification and stability of the structural possibilities of black hole crease sets. In this work we revisit the strategy adopted by Gadioux \& Reall~\cite{GadRea23} for such a classification in the setting of singularity theory in contact geometry. Specifically, in such contact geometry setting, the event horizon is identified as a component (not connected to null infinity) of a so-called  ``BigFront''. The characterization of BigFronts as Legendrian projections of Legendrian submanifolds permits to classify the crease sets and ``cuspidal sets'' (or caustics in Penrose's terminology) by applying classical results established by V.I. Arnol'd. Here we refine the stability discussion presented in \cite{GadRea23} of that connected component of the crease set that is not causally connected to null infinity and that constitutes the event horizon birth. In addition, we identify the existence of other components of the crease set that lie in the part of the BigFront that is causally connected to null infinity.

\end{abstract}
\section{Introduction}
In 1968, R. Penrose published one of his early works on the formation of black hole event horizons. Specifically, in ~\cite{dewitt1968battelle}, Penrose defined the so-called crossover of a black hole event horizon as the set of points where null geodesics meet and enter into the horizon. The limit points of such set, where null geodesics converge at "caustics," along with the crossover, comprise the crease set of a black hole. In this study, we examine structurally stable crease sets of black hole horizons using the contact geometry framework, a geometric tool closely related to the so-called catastrophe theory~\cite{ArnBTCT}. For the sake of clarity, we will focus on black hole horizons with spherical topology formed during a gravitational collapse process. To study black hole event horizon formation under the terminology used in contact geometry, we study the crease set of an event horizon as the cuspidal edge and self-intersection set of a so-called BigFront.

From the perspective of Huygens' wave propagation theory in $\R^3$, a $t$-equidistant wavefront from a light source $S\subset \mathbb{R}^3$ is the set of points reached by light rays emitted orthogonally from $S$ and traveling an Euclidean distance $t$. In spacetime, the union of the wavefronts equidistant to $S$ is a null hypersurface that shares the same geometrical and structural features of wavefronts, known as BigFront. Light phenomena show that caustics have non-regular points. The latter implying a non-regularity on the wavefronts, called singularities in contact geometry \cite{ArnUri}. The union in spacetime of such singularities is a subset of the BigFront, called the cuspidal edge. One of the most significant insights in the study of caustics and wavefronts was made by V. I. Arnol'd in 1972, when he recognized that the underlying structure of these phenomena is governed by the theory of singularities of differentiable maps. In \cite{Arnold1972}, Arnol'd established a classification of the stable singularities that appear on caustics and wavefronts. By using such classification, we show that there is only one type of stable singularities in the crease set of black hole horizons with spherical topology. 

\section{Contact Geometry in Lorentzian Geometry}
\begin{figure}[h]
	\begin{subfigure}{0.4\textwidth}
		\centering
		\includegraphics[width=8.5cm,height=5cm]{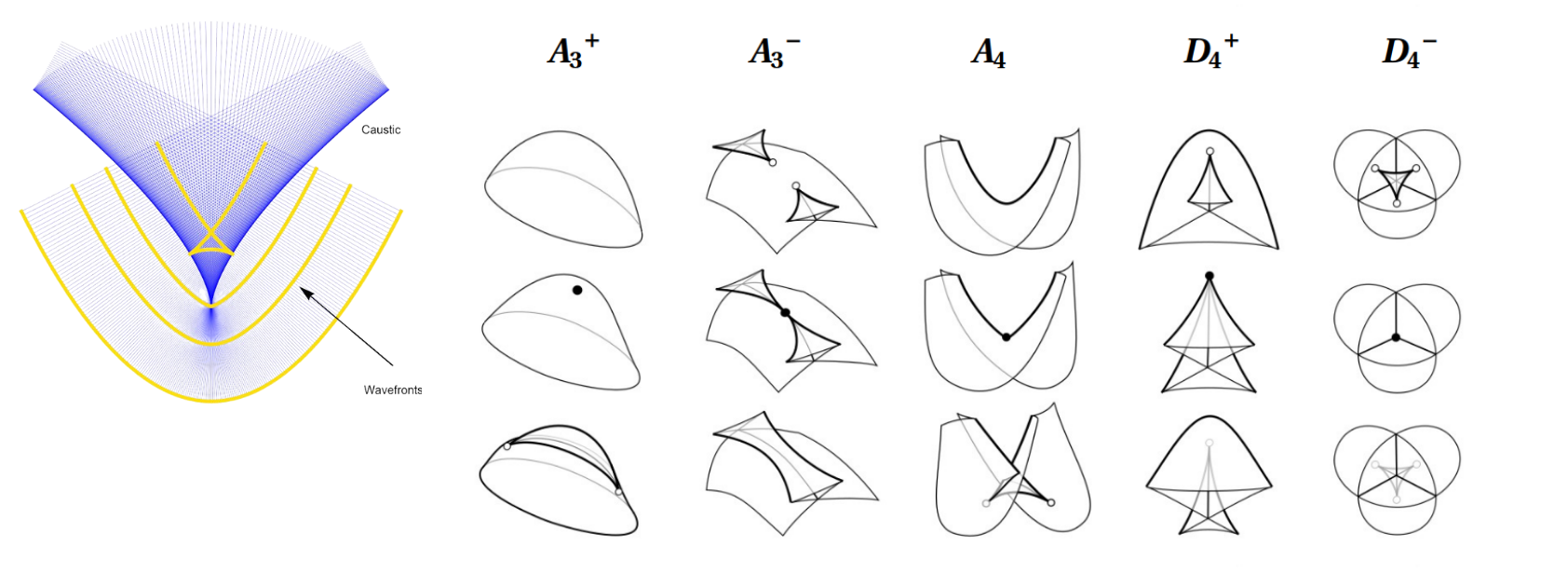}
		\caption{Local classification of singularities in dimension $2$ and $3$. Figure of 2-dimensional front perestroikas taken from \cite{ArnUri}.}
	\end{subfigure}
	\hfil\qquad
	\begin{subfigure}{0.4\textwidth}
		\centering
		\includegraphics[width=6.5cm,height=5cm]{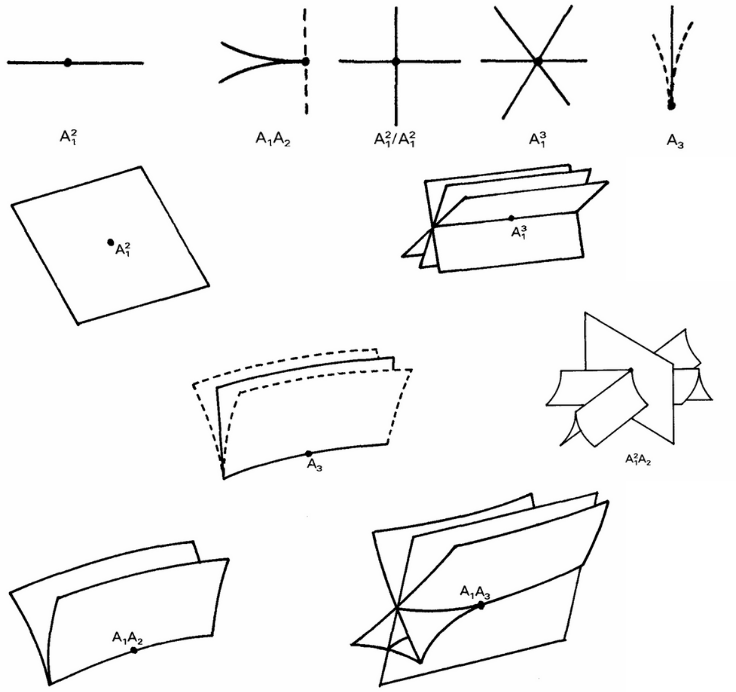}
		\caption{Local classification of self-intersection sets in dimension $2$ and $3$. Figure taken from \cite{GiGi85}.}
	\end{subfigure}	
	\caption{Classification of singularities and self-intersection sets in wavefronts.} \label{F:Classification}
\end{figure}

A contact manifold is an odd-dimensional smooth manifold endowed with a distribution of hyperplanes that are maximally non-integrable. If the hyperplane distribution is given by the kernel of a differentiable 1-form $\alpha$, called contact structure, the non-integrability condition is algebraically expressed as $\alpha\wedge (d\alpha)^n\not=0$. The hyperplanes, called contact planes, are symplectic vector spaces with the restriction of $d\alpha$ to the hyperplanes as symplectic form, that is, $\omega:=d\alpha\vert_{\textit{Ker}\alpha}$ is a non-degenerate skew symmetric two form. 
\begin{definition}
	Let $(\M,g)$ be a $1+n$-dimensional time-oriented spacetime. A contact element based at a point $q\in\M$ is a tangent hyperplane of $T_q\M$. A co-orientation of the contact element is the contact element together with the choice of one of the two half spaces into which it divides the tangent space. A light element based at a point $q\in\M$ is a contact element tangent to the null cone.
\end{definition}
The set of all co-oriented contact elements, denoted as $ST^*\M$, is a contact manifold with a contact structure defined as follows. By writing as $\pi:ST^*\M\rightarrow\M$, $(q,H_q)\mapsto q$, where $H_q$ is a contact element, the map that sends a contact element to its base point, the hyperplane $\Pi_{q,H_q}:=\pi^{-1}_{*,q}(H_q)$ is a contact plane~\cite{ArnoldEQ}. In $ST^*\M$, the set of all light elements based at each point in $\M$ is a hypersurface, called the hypersurface of light elements and denoted by $K$. At a point $Q=(q,H_q)\in K$, the hyperplane $P_Q=\Pi_Q\cap T_QK$ is a $2n-1$-dimensional plane contained in a $2n$ symplectic vector space. The symplectic orthogonal to $P_Q$, defined as 
\begin{equation*}
	l_Q=\{v\in\Pi_Q\ ;  \omega(v,u)=0 \ \forall u\in P_Q \ \}
\end{equation*}
 is a tangent line at the point $Q$. The set of such lines define a vector field whose vector flow is a set of curves called characteristic curves~\cite{ArnoldEQ}. The projection of the characteristic curves to $\M$ via $\pi$ are the null geodesics in $\M$. The latter follows from the next result by computing the characteristic curves in $K$.
\begin{lemma}
	\label{light-elem-deco}
	Let $q\in \M$ and consider a null vector $l\in T_q\M$. The kernel of the 1-form  $\alpha_{q,l}:=g_q(l,\cdot)$ is a tangent hyperplane to the null cone in $T_q\M$. 
\end{lemma}
 Let $N^k\subset ST^*\M$ be a $k$-dimensional smooth manifold with $k\leq n$. The submanifold $N^k$ is said to be integral if $T_QN^k\subset \Pi_Q$ for all $Q\in N^k$. If $k=n$, then $N^n$ is said to be a Legendrian submanifold. A point $Q\in N^k$ is said to be non-characteristic if $l_Q\not\subset T_QK$. 
\begin{definition}
  Let $N^{n-1}\subset K$ be a non-characteristic integral submanifold. The Cauchy problem for the hypersurface of light elements with $N^{k-1}$ as initial condition consist in finding a Legendrian submanifold $L^n\subset K$ containing $N^{k-1}$. Such Legendrian submanifold, $L^n$, is called a \textit{null Legendrian submanifold}. 
\end{definition}
\begin{theorem}\label{T:Exi-Uniq}{\textnormal{\cite{ArnoldEQ}}}
	Let $Q\in N^{n-1}$ be a non-characteristic point. Then, there exists a neighborhood $U$ of $Q$ where the Cauchy problem for the hypersurface $K$ with initial contidion $N$ exists and is unique. 
\end{theorem}
Let $L\xhookrightarrow{\iota}ST^*\M$ be an immersed Legendrian submanifold. The map $\pi\circ\iota:L\rightarrow\M$ is called a Legendrian map and the image $\pi\circ\iota(L)$ is called a  front. If $L$ is a null Legendrian submanifold solution to the Cauchy problem for $K$ with $N$ as initial condition, the image of the Legendrian map, called BigFront, is a null hypersurface. The projection of non-characteristic integral sections of $L$ to $\M$ via $\pi$ are called wavefronts. The projection of the initial condition $N$ to $\M$ via $\pi$ is the light source $S$.
\begin{lemma}\label{L:CauchysolN}
	Let $\Gamma^{1+n-k}$ be a spacelike submanifold. Then, $\Gamma^{1+n-k}$ is lifted to a non-characteristic initial integral submanifold $N^{n-1}$ in the hypersurface of light elements. Moreover, the solution to the Cauchy problem for the hypersurface of light elements with initial condition $N^{n-1}$ exists and is unique. 
\end{lemma}
The proof of Lemma \ref{L:CauchysolN} is a consequence of Theorem \ref{T:Exi-Uniq} by taking $N$ as the submanifold made of light elements tangent to $\Gamma$. Lemma \ref{L:CauchysolN} implies that for a given spacelike surface, interpreted as a light source or an initial wavefront, its BigFront is a null hypersurface that contains the dynamical evolution of the initial wavefront propagating along null geodesics orthogonal to $\Gamma$. The set of critical values of a Legendrian map is the set of points where the front is non-smooth. In singularity theory, the generic and stable singularities are known and are classified for any smooth manifold of dimension less or equal than 6. In a $4$-dimensional spacetime, the singularities of Legendrian maps are given by the following result.
\begin{theorem}\textnormal{(Arnol'd,\cite{Arnold1972})}
	The only simple and stable singularities in wavefronts for dimension 2 are cusps and transverse self-intersections. In dimension 3, the local singularities are showed in Fig. \ref{F:Classification}.
\end{theorem} 
Consequently,  the local shape of the simple stable singularities in dimension 2 and 3 allows to describe the local shape of the self-intersection set in the wavefronts. 
\begin{theorem}\textnormal{(Giblin et al., \cite{GiGi85})}
	The local shape of the self-intersection sets in dimension 2 and 3 is showed in Fig. \ref{F:Classification}.
\end{theorem}
By definition, the self-intersection set is the set of points where two sheets of a given surface meet. If the given surface is a BigFront, then it is possible to know the local shape of its self-intersection set by analyzing the self-intersection set of the wavefronts that forms the BigFront (also known as symmetry sets. See~\cite{GiGi85,Meneses2025} for a formal study). The union in spacetime of the self-intersection set of each wavefront is the self-intersection set of the BigFront, whose boundary consists of points of the cuspidal set. 
\section{Black Hole event horizons as BigFronts}
As a toy model, let us consider a spacetime where the dynamics is subject to physical energy conditions for the stress-energy tensor so that a mass distribution collapses, thereby ensuring the existence of a black hole horizon with a crease set (\cite{AdlBj0Che05,Pen65,OppSny39,Pen72}). Let us consider now that at late times, a spacelike section of the horizon is topologically a sphere. By taking such section as initial wavefront and by following the null geodesics orthogonal to such section to the past, the event horizon is a BigFront with a self-intersection and a cuspidal set~\cite{Siino98}. By Penrose Theorems, the crease set is formed by crossover point and caustics \cite{dewitt1968battelle}. In the BigFront perspective, the crossover is a connected component (the first one when coming from the past) of the self-intersection set of the BigFront. The caustic corresponds to the cuspidal set.
\begin{theorem}\textnormal{(Meneses-Rojas, \cite{Meneses2025})}\label{T:Birth}
	Let $W$ be the null hypersurface generated by the null geodesics orthogonal to the spacelike 2-dimensional surface $\sigma_2\subset \Ho$, where $\Ho\subset W$ is the event horizon resulting from a gravitational collapse process. Let $\sigma_1$ be the spacelike section of $W$ such that it is no longer in the event horizon. Assume the topology of $\sigma_2$ to be a sphere. Then, $W$ is the BigFront of a Legendrian map $L\subset ST^*\M\rightarrow\M$ where $L$ is a null Legendrian submanifold such that in the perestroikas from $\sigma_2$ to $\sigma_1$ (the wavefronts) there is a cuspidal and a self-intersection set on the BigFront. Moreover, a component of the self-intersection of $W$ set splits the BigFront into two components: the non-visible (the event horizon $\Ho$) and the component causally connected to $\mathscr{I}^+$.
\end{theorem}
The proof of Theorem \ref{T:Birth} follows by rewriting the results of Penrose in the contact geometry framework and recognizing that the self-intersection set of a BigFront suits with the description of the crossover set. The results of Giblin et al.~\cite{GiGi85,GiKi2000} in this scenario allows to give a precise description of the self-intersection set and the cuspidal set of event horizons "births" as a result of a gravitational collapse process. 
\begin{theorem}\textnormal{(Meneses-Rojas, \cite{Meneses2025})}\label{T:Uniq}
	The stable cuspidal set of the event horizon is of type $A_3$ in Fig. \ref{F:Classification}.
\end{theorem}
In conclusion, by using the geometric settings of singularities in contact geometry, it is possible to re-interpret the event horizon as a component of a BigFront of a Legendrian map. Thus, the crease set is naturally analyzed as a connected component  of the self-intersection set of the entire BigFront. Moreover, the local shape of the singularities is known. If in addition, by Theorem \ref{T:Uniq}, there is only one possible type of singularity allowed in the event horizon, then it is possible to study the caustic associated to such singularity (e.g. diffraction at a caustics \cite{Berry01011976}). 

\section{Acknowledgments}
The author is grateful to his PhD advisors Jose Luis Jaramillo and Ricardo Uribe-Vargas for proposing the problem, the discussions and ideas shared that came along this work.
\bibliography{iopart-num}

\providecommand{\newblock}{}
\begin{thebibliography}{10}
\expandafter\ifx\csname url\endcsname\relax
  \def\url#1{{\tt #1}}\fi
\expandafter\ifx\csname urlprefix\endcsname\relax\def\urlprefix{URL }\fi
\providecommand{\eprint}[2][]{\url{#2}}

\bibitem{GadRea23}
Gadioux M and Reall H~S 2023 {\em Phys. Rev. D\/} {\bf 108}(8) 084021
  \urlprefix\url{https://link.aps.org/doi/10.1103/PhysRevD.108.084021}

\bibitem{dewitt1968battelle}
DeWitt-Morette C, Wheeler J and Institute B~M 1968 {\em Battelle Rencontres:
  1967 Lectures in Mathematics and Physics\/} (Benjamin)
  \urlprefix\url{https://books.google.fr/books?id=u2w-AQAAIAAJ}

\bibitem{ArnBTCT}
Arnold V~I 1994 {\em Bifurcation theory and catastrophe theory / V.I. Arnol'd
  (ed.).\/} Dynamical systems ; 5 (Berlin ;: Springer-Verlag) ISBN 3540181733

\bibitem{ArnUri}
Arnold V and Uribe-Vargas R {\em Geometry\/} (Unpublished book)

\bibitem{Arnold1972}
Arnol'd V~I 1972 {\em Functional Analysis and Its Applications\/} {\bf 6}
  254--272 ISSN 1573-8485 \urlprefix\url{https://doi.org/10.1007/BF01077644}

\bibitem{ArnoldEQ}
Arnol'd V~I 1980 {\em Chapitres supplémentaires de la théorie des équations
  différentielles ordinaires\/} (Éditions MiR. Moscou)

\bibitem{GiGi85}
Bruce J~W, Giblin P~J and Gibson C~G 1985 {\em Proceedings of the Royal Society
  of Edinburgh: Section A Mathematics\/} {\bf 101} 163–186

\bibitem{Meneses2025}
Meneses~Rojas O 2025 {\em Singularity Theory and Gravitation: Caustics and
  Wavefronts as probes into Black Holes\/} Ph.D. thesis

\bibitem{AdlBj0Che05}
Adler R~J, Bjorken J~D, Chen P and Liu J~S 2005 {\em American Journal of
  Physics\/} {\bf 73} 1148--1159 ISSN 0002-9505 (\textit{Preprint}
  \eprint{https://pubs.aip.org/aapt/ajp/article-pdf/73/12/1148/13083948/1148\_1\_online.pdf})
  \urlprefix\url{https://doi.org/10.1119/1.2117187}

\bibitem{Pen65}
Penrose R 1965  {\bf 14} 57

\bibitem{OppSny39}
Oppenheimer J~R and Snyder H 1939 {\em Phys. Rev.\/} {\bf 56}(5) 455--459
  \urlprefix\url{https://link.aps.org/doi/10.1103/PhysRev.56.455}

\bibitem{Pen72}
Penrose R 1972 {\em Techniques of Differential Topology in Relativity\/}
  (Society for Industrial and Applied Mathematics) (\textit{Preprint}
  \eprint{https://epubs.siam.org/doi/pdf/10.1137/1.9781611970609})
  \urlprefix\url{https://epubs.siam.org/doi/abs/10.1137/1.9781611970609}

\bibitem{Siino98}
Siino M 1998 {\em Progress of Theoretical Physics\/} {\bf 99} 1--32 ISSN
  0033-068X (\textit{Preprint}
  \eprint{https://academic.oup.com/ptp/article-pdf/99/1/1/5428199/99-1-1.pdf})
  \urlprefix\url{https://doi.org/10.1143/PTP.99.1}

\bibitem{GiKi2000}
Giblin P and Kimia B 2000 A formal classification of 3d medial axis points and
  their local geometry vol~26 pp 566--573 vol.1 ISBN 0-7695-0662-3

\bibitem{Berry01011976}
Berry M 1976 {\em Advances in Physics\/} {\bf 25} 1--26 (\textit{Preprint}
  \eprint{https://doi.org/10.1080/00018737600101342})
  \urlprefix\url{https://doi.org/10.1080/00018737600101342}

\end{thebibliography}

\end{document}